\documentclass[graphicx,preprint,11pt]{aastex}

\usepackage{color}

\lefthead{}
\righthead{}

\begin{document}

\title{Fossilized condensation lines in the Solar System protoplanetary disk} 

\author{\textbf{A. Morbidelli$^{(1)}$, B. Bitsch$^{(2)}$, A. Crida$^{(1,4)}$, M. Gounelle$^{(3,4)}$, T. Guillot$^{(1)}$, S. Jacobson$^{(1,5)}$, A. Johansen$^{(2)}$, M. Lambrechts$^{(1)}$, E. Lega$^{(1)}$}\\  
(1) Laboratoire Lagrange, UMR7293, Universit\'e C\^ote d'Azur,
  CNRS, Observatoire de la C\^ote d'Azur. Boulevard de l'Observatoire,
  06304 Nice Cedex 4, France. (Email: morby@oca.eu / Fax:
  +33-4-92003118) \\
(2) Lund Observatory, Department of Astronomy and Theoretical Physics, Lund University, Box 43, 22100, Lund, Sweden  \\ 
(3) IMPMC, Mus\'eum national d'histoire naturelle, Sorbonne Universit\'es, CNRS, UPMC \& IRD, 57 rue Cuvier 75005 Paris, France \\
(4) Institut Universitaire de France
103 bd Saint Michel, 75005 Paris, France\\
(5) Bayerisches Geoinstitut, University of Bayreuth, D-95440 Bayreuth, Germany
} 

\begin{abstract}
The terrestrial planets and the asteroids dominant in the inner asteroid belt are water poor. However, in the protoplanetary disk the temperature should have decreased below water-condensation level well before the disk was photo-evaporated. Thus, the global water depletion of the inner Solar System is puzling. We show that, even if the inner disk becomes cold, there cannot be direct condensation of water. This is because the snowline moves towards the Sun more slowly than the gas itself. Thus the gas in the vicinity of the snowline always comes from farther out, where it should have already condensed, and therefore it should be dry. The appearance of ice in a range of heliocentric distances swept by the snowline can only be due to the radial drift of icy particles from the outer disk. However, if a planet with a mass larger than 20 Earth mass is present, the radial drift of particles is interrupted, because such a planet gives the disk a super-Keplerian rotation just outside of its own orbit. From this result, we propose that the precursor of Jupiter achieved this threshold mass when the snowline was still around 3 AU. This effectively fossilized the snowline at that location. In fact, even if it cooled later, the disk inside of Jupiter's orbit remained ice-depleted because the flow of icy particles from the outer system was intercepted by the planet. This scenario predicts that planetary systems without giant planets should be much more rich in water in their inner regions than our system. We also show that the inner edge of the planetesimal disk at 0.7~AU, required in terrestrial planet formation models to explain the small mass of Mercury and the absence of planets inside of its orbit, could be due to the silicate condensation line, fossilized at the end of the phase of streaming instability that generated the planetesimal seeds. Thus, when the disk cooled, silicate particles started to drift inwards of 0.7~AU without being sublimated, but they could not be accreted by any pre-existing planetesimals. 
\end{abstract}

\section{Introduction}

The chemical structure of a protoplanetary disk is characterized by a condensation front for each chemical species. It marks the boundary beyond which the temperature is low enough to allow the condensation of the considered species, given its local partial pressure of gas. If one assumes that the disk is vertically isothermal and neglects pressure effects, the condensation front is a vertical straight-line in $(r,z)$ space. This is the reason for the wide-spread use of the term ``condensation line''. However, the vertical isothermal approximation is in many cases a poor proxy for the thermal structure of the disk (see below), so that in reality the condensation ``line'' is a curve in $(r,z)$ space, {like any other isothermal curve (Isella and Natta, 2005)}.  

Probably the most important condensation line is that for water, also called the ice-line or the snowline. In the Solar System water accounts for about 50\% of the mass of all condensable species (Lodders, 2003). The fact that the inner Solar System objects (terrestrial planets, asteroids of the inner main belt) are water poor, whereas the outer Solar System objects (the primitive asteroids in the outer belt, most satellites of the giant planets and presumably the giant planets cores, the Kuiper belt objects and the comets) are water rich, argues for the importance of the snowline in dividing the protoplanetary disk in two chemically distinct regions.


Thus, modeling the thermal structure of the disk has been the subject of a number of papers. 
There are two major processes generating heat: viscous friction and stellar irradiation.
Chiang and Goldreich (1997), Dullemond et al. (2001, 2002) and Dullemond (2002) neglected viscous heating and considered only stellar irradiation of passive disks. They also assumed a constant opacity (i.e. independent of temperature). Chiang and Goldreich demonstrated the flared structure of a protoplanetary disk while the Dullemond papers stressed the presence of a puffed-up rim due to the face illumination of the disk's inner edge. This rim casts a shadow onto the disk, until the flared structure brings the outer disk back into illumination.  Hueso and Guillot (2005), Davis (2005), Garaud and Lin (2007), Oka et al. (2011), Bitsch et al. (2014a, 2015) and Bailli\'e et al. (2015) considered viscous heating also and introduced temperature dependent opacities with increasingly sophisticated prescriptions. They demonstrated that viscous heating dominates in the inner part of the disk for $\dot{M}>10^{-10} M_\odot/y$ (Oka et al., 2011), where $\dot{M}$ is the radial mass-flux of gas (also known as the stellar accretion rate) sustained by the viscous transport in the disk. In the most sophisticated models, the aspect ratio of the disk is grossly independent of radius in the region where the viscous heating dominates, although bumps and dips exist (with the associated shadows) due to temperature-dependent transitions in the opacity law (Bitsch et al., 2014a, 2015). The temperature first decreases with increasing distance from the midplane, then increases again due to the stellar irradiation of the surface layer. The outer part of the disk is dominated by stellar irradiation and is flared as predicted earlier; the temperature in that region is basically constant with height near the mid-plane and then increases approaching the disk's surface. 

As a consequence of this complex disk structure, the snowline is a curve in the $(r,z)$ plane (see for instance Fig. 4 of Oka et al., 2011). On the midplane, the location of the snowline is at about 3~AU when the accretion rate in the disk is $\dot{M}=3$--$10\times 10^{-8} M_\odot/y$. When the accretion rate drops to $5$--$10\times 10^{-9}M_\odot/y$ the snowline on the midplane has moved to 1~AU (Hueso and Guillot, 2005; Davis, 2005; Garaud and Lin, 2007; Oka et al., 2011; Bailli\'e et al., 2015; Bitsch et al., 2015). Please notice that a disk should not disappear before that the accretion rate decreases to $\dot{M}\lesssim 10^{-9} M_\odot/y$ (Alexander et al., 2014). The exact value of the accretion rate for a given snowline location depends on the disk model (1+1D as in the first four references or 2D as in the last one) and on the assumed dust/gas ratio and viscosity but does not change dramatically from one case to the other for reasonable parameters, as we will see below (eq.~\ref{temp}).

The stellar accretion rate as a function of age can be inferred from observations. Hartmann et al. (1998) found that on average $\dot{M}=10^{-8} M_\odot/y$ at 1~My and $\dot{M}=1$--5$\times 10^{-9}M_\odot/y$ at 3~My. The accretion rate data, however, appear dispersed by more than an order of magnitude for any given age (possibly because of uncertainties in the measurements of the accretion rates and in the estimates of the stellar ages, but nevertheless there should be a real dispersion of accretion rates in nature). In some cases, stars of 3-4~My may still have an accretion rate of $10^{-8} M_\odot/y$  (Hartmann et al., 1998; Manara et al., 2013). 

The Solar System objects provide important constraints on the evolution of the disk chemistry as a function of time. Chondritic asteroids are made of chondrules. The ages of chondrules span the $\sim 3$ My period after the formation of the first solids, namely the calcium-aluminum inclusions (CAIs; Villeneuve et al., 2009; Connelly et al., 2012; Bollard et al., 2014; Luu et al., 2015). The measure of the age of individual chondrules can change depending on which radioactive clock is used, but the result that chondrule formation is protracted for $\sim 3$~My seems robust. Obviously, the chondritic parent bodies could not form before the chondrules. Hence, we can conclude that they formed (or continued to accrete until; Johansen et al., 2015) 3-4 My after CAIs. 

At 3~My (typically $\dot{M}=1$--5$\times 10^{-9}M_\odot/y$) the snowline should have been much closer to the Sun than the inner edge of the asteroid belt (the main reservoir of chondritic parent bodies). Nevertheless, ordinary and enstatite chondrites contain very little water (Robert, 2003). {Some water alteration can be found in ordinary chondrites (Baker et al., 2003) as  well as clays produced by the effect of water (Alexander et al. 1989). Despite these observations, it seems very unlikely that the parent bodies of these meteorites ever contained $\sim$~50\% of water by mass, as expected for a condensed gas of solar composition (Lodders, 2003). 

One could think that our protoplanetary disk was one of the exceptional cases still showing stellar accretion $\gtrsim 10^{-8} M_\odot/y$ at $\sim 3$~My. However, this would not solve the problem. In this case the disk would have just lasted longer, while still decaying in mass and cooling. In fact, the photo-evaporation process is efficient in removing the disk only when the accretion rate drops at $\lesssim 10^{-9}M_\odot/y$ (see Fig. 4 of Alexander et al., 2014). Thus, even if the chondritic parent bodies had formed in a warm disk, they should have accreted a significant amount of icy particles when, later on, the temperature decreased below the water condensation threshold, but before the disk disappeared.

The Earth provides a similar example. Before the disk disappears ($\dot{M}\sim  10^{-9}M_\odot/y$), the snowline is well inside 1~AU (Oka et al., 2011). Thus, one could expect that plenty of ice-rich planetesimals formed in the terrestrial region and our planet accreted a substantial fraction of water by mass. Instead, the Earth contains no more than $\sim 0.1$\% of water by mass (Marty, 2012). The water budget of the Earth is perfectly consistent with the Earth accreting most of its mass from local, dry planetesimals and just a few percent of an Earth mass from primitive planetesimals coming from the outer asteroid belt, as shown by dynamical models (Morbidelli et al., 2000; Raymond et al.,  2004, 2006, 2007; O'Brien et al., 2006, 2014).  Why water is not substantially more abundant on Earth is known as the {\it snowline problem}, first pointed out clearly by Oka et al. (2011). Water is not an isolated case in this respect. The Earth is depleted in all volatile elements (for lithophile volatile elements the depletion progressively increases with decreasing condensation temperature; McDonough and Sun, 1995). Albarede (2009), using isotopic arguments, demonstrated that this depletion was not caused by the loss of volatiles during the thermal evolution of the planet, but is due to their reduced accretion relative to solar abundances. {Furthermore, a significant accretion of oxidized material would have led to an Earth with different chemical properties (Rubie et al., 2015).} Mars is also a water-poor planet, with only 70--300ppm of water by mass (McCubbin et al., 2012).

Thus, it seems that the water and, more generally, the volatile budget of Solar System bodies reflects the location of the snowline at a time different from that at which the bodies formed. Interestingly and never pointed out before, the situation may be identical for refractory elements. In fact, a growing body of modeling work (Hansen, 2009; Walsh et al., 2011; Jacobson and Morbidelli, 2014) suggests that the disk of planetesimals that formed the terrestrial planets had an inner edge at about 0.7~AU. This edge is required in order to produce a planet of small mass like Mercury (Hansen, 2009). On the midplane, a distance of 0.7~AU corresponds to the condensation line for silicates (condensation temperature $\sim 1300$~K) for a disk with accretion rate $\dot{M}\sim 1.5\times 10^{-7} M_\odot/y$, typical of an early disk. Inside this location, it is therefore unlikely that objects could form near time zero. The inner edge of the planetesimal disk at 0.7~AU then seems to imply that, for some unknown reason, objects could not form there even later on, despite the local disk's temperature should have dropped well below the value for the condensation of silicates. Clearly, this argument is more speculative than those reported above for the snowline, but it is suggestive that the snowline problem is common to all chemical species. It seems to indicate that the structure of the inner Solar System carries the {\it fossilized} imprint of the location that the condensation lines had at an early stage of the disk, rather than at a later time, more characteristic of planetesimal and planet formation; hence the title of this Note. Interestingly, if this analogy between the silicate condensation line and the snowline is correct, the time of fossilization of these two lines would be different (the former corresponding to the time when $\dot{M}\sim 1.5\times 10^{-7} M_\odot/y$, the latter when $\dot{M}\sim 3\times 10^{-8} M_\odot/y$).

The goal of this Note is to discuss how this might be understood. This Note will not present new sophisticated calculations, but simply put together results already published in the literature and connect them to propose some considerations, to our knowledge never presented before, that may explain the fossilization of the condensation lines, with focus on the snowline and the silicate line. 

Below, we start in Sect.~2 with a brief review of scenarios proposed so far to solve the snowline and the 0.7 AU disk edge problems. In Sect. 3 we discuss gas radial motion, the radial displacement of the condensation lines and the radial drift of solid particles. This will allow us to conclude that the direct condensation of gas is not the main process occurring when the temperature decreases, but instead it is the radial drift of particles from the outer disk that can repopulate the inner disk of condensed species. With these premises, in Sect. 4 we focus on the snowline, and discuss mechanisms for preventing or reducing the flow of icy particles, so to keep the Solar System deficient in ice inside $\sim 3$~AU even when the temperature in that region dropped below the ice-condensation threshold. In Section~5 we link the inner edge of the planetesimal disk to the original location of the silicate condensation line and we  attempt to explain why no planetesimals formed inside this distance when the temperature dropped. A wrap-up will follow in Sect. 6 and an Appendix on planet migration in Sect. 7.

\section{Previous models}
\label{old}

The condensation line problem is a subject only partially explored.  For the snowline problem, Martin and Livio (2012, 2013) proposed that the dead zone of the protoplanetary disk piled up enough gas to become gravitationally unstable. The turbulence driven by self-gravity increased the temperature of the outer parts of the dead zone and thus the snowline could not come within 3~AU, i.e. it remained much farther from the star than it would in a normal viscously evolving disk. This model, however, has some drawbacks. First, it predicts an icy region inside of the Earth's orbit, so that Venus and Mercury should have formed as icy worlds. {Second, from the modeling standpoint, the surface density ratio between the deadzone and the active zone of the disk is inversely proportional to the viscosity ratio only in 1D models of the disk. In 2D $(r,z)$ models (Bitsch et al., 2014b) the relationship between density and viscosity is non-trivial because the gas can flow in the surface layer of the disk. Thus, the deadzone may not become gravitationally unstable.}

Hubbard and Ebel (2014) addressed the deficiency of the Earth in lithophile volatile elements. They proposed that grains in the protoplanetary disk are originally very porous. Thus, they are well coupled with the gas and distributed quite uniformly along the vertical direction. The FU-Orionis events, that our Sun presumably experienced like most young stars, would have heated above sublimation temperature the grains at the surface of the disk. Then, the grains would have recondensed, losing the volatile counterpart and acquiring a much less porous structure and a higher density. These reprocessed grains would have preferentially sedimented onto the disk's midplane, featuring the major reservoir of solids for the accretion of planetesimals and the planets. Planetesimals and planets would therefore have accreted predominantly from volatile depleted dust, even though the midplane temperature was low. This model is appealing, but has the problem that the phase of FU-Orionis activity of a star lasts typically much less than the disk's lifetime. Thus eventually the devolatilazation of the grains would stop and the planetesimals and planets would keep growing from volatile-rich grains. Also, it neglects the radial drift of icy particles on the mid-plane from the outer disk. 

Concerning the inner edge of the planetesimal disk at 0.7 AU, an explanation can be found in Ida and Lin (2008). The authors pointed out that the timescale for runaway growth of planetary embryos decreases with heliocentric distance. Because the radial migration speed of embryos is propotional to their mass (Tanaka et al., 2002), the innermost embryos are lost into the star and are not replaced at the same rate by embryos migrating inward from farther out. This produces an effective inner edge in the solid mass of the disk, that recedes from the Sun as time progresses (see Fig. 2 of Ida and Lin, 2008).  The major issue here is whether planets and embryos can really be lost into the star. The observation of extrasolar planets has revealed the existence of many ``hot'' planets, with orbital periods of a few days. Clearly, these planets would be rare if there had existed no stopping mechanism to their inward migration, probably due to the existence of an inner edge of the protoplanetary disk where the Keplerian period is equal to the star's rotation period ({Koenigl, 1991; Lin et al., 1996}), acting like a planet-trap (Masset et al., 2006). The presence of planet-trap  would change completely the picture presented in Ida and Lin (2008) {(see for instance Cossou et al., 2014).}

More recently, Batygin and Laughlin (2015) and Volk and Gladman (2015) proposed that the the Solar System formed super-Earths inside of 0.7~AU, but these planets were lost, leaving behind only the ``edge'' inferred by terrestrial planet formation models. In Batygin and Laughlin (2015), the super-Earths are pushed into the Sun by small planetesimals drifting by gas-drag towards the Sun and captured in mean motion resonances. Again, we are faced with the issue {of the probable presence of a planet-trap at the inner edge of the protoplanetary disk.  With a planet-trap,} the super-Earths would probably not have been removed despite of the planetesimals push. In Volk and Gladman (2015), instead, the super-Earths become unstable and { start to collide with each other at velocities large enough for these collisions to be erosive. There is no explicit modeling, however, of the evolution of the system under these erosive collisions. We expect that the debris generated in the first erosive collisions would exert dynamical friction on the planets and help them achive a new, stable configuration (see for instance Chambers, 2013). Thus, we think it is unlikely that a system of super-Earths might disappear in this way.} 

From this state-of-art literature analysis it appears that the condensation line problem is still open. Thus, we believe  that it is interesting to resume the discussion and approach the problem globally, i.e. addressing the general issue of the ``fossilization'' of condensation lines at locations corresponding to some ``early'' times in the disk's life. 

\section{Relevant radial velocities}
\label{radvels}

In this section we review the radial velocities of the gas, of the condensation lines and of solid particles. This will be important to understand how a portion of the disk gets enriched in condensed elements as the disk evolves and cools, and it will give hints on how a region could remain depleted in a chemical species even when the temperature drops beyond its corresponding condensation value. 
  
The seminal work for the viscous evolution of a circumstellar disk is Lynden-Bell and Pringle (1974). We consider the disk described in their section 3.2, which can be considered as the archetype of any protoplanetary disk, which accretes onto the star while spreading in the radial direction under the effect of viscous transport. The viscosity $\nu$ is assumed to be constant with radius in Lynden-Bell and Pringle's work, but the results we will obtain below are general for a viscously evolving disk, even with more realistic prescriptions for the viscosity (whenever we need to evaluate the viscosity, we will then adopt the $\alpha$ prescription of Shakura and Sunyaev, 1973).

According to eq. (18') of Lynden-Bell and Pringle (1974), the radial velocity of the gas is
\begin{equation}
u_r=-{{3}\over{2}} {{\nu}\over{r}} \left[1-{{4a(GMr)^2}\over{\tau}}\right]\ ,
\label{ur}
\end{equation}
where $G$ is the gravitational constant, $M$ is the mass of the central star, $a$ is a parameter describing how sharp is initially the outer edge of the disk and $\tau$ is a normalized time, defined as
\begin{equation}
\tau=\beta\nu t +1\ ,
\end{equation}
where $t$ is the natural time and $\beta=12(GM)^2 a$. 
Still according to the same paper, the surface density of the gas evolves as:
\begin{equation}
\Sigma={{C \tau^{-5/4}}\over{3\pi\nu}}\exp\left[-{{a(GMr)^2}\over{\tau}}\right]\ ,
\label{sigma}
\end{equation}
where $C$ is a parameter related to the peak value of $\Sigma$ at $r=0$ and $t=0$. The disk described by this equation spreads with time (the term $a(GMr)^2/\tau$ becoming smaller and smaller with time), while the peak density declines as $\tau^{-5/4}$. The motion of the gas is inwards for $r<r_0\equiv\sqrt{\tau/(4a)}/(GM)$ and outwards beyond $r_0$, which itself moves outwards as $r_0\propto \sqrt{t}$. 

We now focus on the inner part of the disk, where $r<<r_0$. In this region we can approximate $a(GMr)^2/\tau$ with $0$ and therefore the equations for the radial velocity of the gas and the density become:
\begin{equation}
u_r=-{{3}\over{2}} {{\nu}\over{r}}\ ,
\label{urin}
\end{equation}
\begin{equation}
\Sigma={{C \tau^{-5/4}}\over{3\pi\nu}}\ .
\label{sigmain}
\end{equation}
Thus, the stellar accretion rate is 
\begin{equation}
\dot{M}=-2\pi r \Sigma u_r(r)= 3\pi\nu\Sigma=C\tau^{-5/4}\ .
\label{dotM}
\end{equation}
 That is, the accretion rate in the inner part of an accretion disk is independent of radius. Eq.~(\ref{urin}) gives the radial velocity of the gas, i.e. the first of the expressions we are interested in. Notice that Takeuchi and Lin (2002) found that, in a three dimensional disk, the radial motion of the gas can be outwards in the midplane and inwards at some height in the disk. Nevertheless the global flow of gas is inwards (the inward flow carries more mass than the outwards flow). The velocity $u_r$ in (\ref{urin}) can be considered as the radial speed averaged along the vertical direction and ponderated by the mass flow. For our considerations below we can consider this average speed, without worrying about the meridional circulation of the gas.  

We now compute the speed at which a condensation line moves inwards in this evolving disk. Neglecting stellar irradiation (which is dominant only in the outer part of the disk; Oka et al., 2011, Bitsch et al., 2015), the temperature on the midplane of the disk can be obtained by equating viscous heating (Bitsch et al., 2013):
\begin{equation}
Q^+= 2\pi r \delta r {{9}\over{8}} \Sigma \nu \Omega^2\ ,
\label{Qplus}
\end{equation}
with radiative cooling:
\begin{equation}
Q^-= 4\pi r \delta r \sigma T^4 /(\kappa \Sigma) \ ,
\label{Qminus}
\end{equation}
where $\Omega=\sqrt{GM/r^3}$ is the orbital frequency, $\sigma$ is Boltzman constant, $\kappa$ is the opacity (here assumed independent of radius and time, for simplicity), $T$ is the temperature and $\delta r$ is the radial width of the considered annulus.
Thus, the expression for the temperature is:
\begin{equation}
T=A[\kappa\nu\Sigma^2]^{1/4} r^{-3/4}=A[\kappa\Sigma\dot{M}/(3\pi)]^{1/4} r^{-3/4}  \ ,
\label{temp}
\end{equation}
where $A=[9GM/(16\sigma)]^{1/4}$. So, the temperature changes with time (through $\Sigma$ and $\dot{M}$) and with radius. Eq.~\ref{temp} also implies that, for a given value of $\dot{M}$, $T$ is weakly dependent (i.e. to the $1/4$ power) on the product $\kappa\Sigma$, namely on the remaining disk parameters. This is why we can link the location of a given condensation line with the disk's accretion rate with small uncertainty. 

The derivatives of the temperature with respect to radius and time are:
\begin{equation}
{{\rm d}T\over{{\rm d}r}}={\partial T\over{\partial r}}=-{{3}\over{4}} A [\kappa\nu\Sigma^2]^{1/4} r^{-7/4} \ ,
\label{dTdr}
\end{equation}
\begin{eqnarray}
{{\rm d}T\over{{\rm d}t}}={\partial T\over{\partial t}}&=&{{1}\over{2}} A[\kappa\nu]^{1/4} r^{-3/4}\Sigma^{-1/2}{{\rm d}\Sigma\over{{\rm d}t}}\cr
&=& -{{5}\over{8}}A[\kappa\nu]^{1/4} r^{-3/4}\Sigma^{-1/2}{{C\tau^{-9/4}}\over{3\pi\nu}}{{\rm d}\tau\over{{\rm d}t}}\cr
&=& -{{5}\over{8}}A[\kappa\nu]^{1/4} r^{-3/4}\Sigma^{-1/2}{{C\tau^{-9/4}\beta}\over{3\pi}}
\end{eqnarray}
(in the derivation of the equations above, please remember that we assumed that $\kappa$ and $\nu$ are constant in time and space and we derived in Eq.~\ref{sigmain} that, at equilibrium, $\Sigma$ in the inner part of the disk is independent of radius, {so that the total derivatives of $T$ are equal to its partial derivatives)}.

Therefore, assuming that the location of a condensation line just depends on temperature (i.e. neglecting the effect of vapor partial pressure), the speed at which a condensation line moves inwards (which is the second expression we are looking for) is:
\begin{equation}
v_r^{cond}=-{{\rm d}r\over{{\rm d}T}} {{\rm d}T\over{{\rm d}t}}=-{{5}\over{6}}{{r}\over{\Sigma}}{{C\tau^{-9/4}\beta}\over{3\pi}}
\label{snowveltmp}
\end{equation}
which, using (\ref{sigmain}) and approximating $\tau$ with $\beta\nu t$, gives:
\begin{equation}
v_r^{cond}=-{{5}\over{6}}{{r}\over{t}}\ .
\label{snowvel}
\end{equation}

By comparing (\ref{snowvel}) with (\ref{urin}) we find that $u_r>v_r^{cond}$ for 
\begin{equation}
t>{{5}\over{9}}{{r^2}\over{\nu}}\ .
\label{condition}
\end{equation}

The inequality (\ref{condition}) implies that, after approximately half a viscous timescale $t_\nu\equiv r^2/\nu$, the radial motion of the gas is faster than the displacement of a given condensation line. The lifetime of a disk is typically several viscous timescales for the inner region. In fact, Hartmann et al. (1998) found that the time-decay of the accretion rate on stars implies that, if one adopts an $\alpha$-prescription for the viscosity (i.e. $\nu=\alpha H^2\Omega$;  Shakura and Sunyaev,  1973), the value of the coefficient $\alpha$ is 0.001--0.01. At 3~AU, assuming a typical aspect ratio of 5\%, the viscous timescale is $t_\nu=3\times 10^4$--$3\times 10^5$~y; at 0.7~AU $t_\nu$ is about 10 times shorter. Both values are considerably shorter than the typical disk's lifetime of a few My. Thus, for the regions and timescales we are interested in (either the asteroid belt at $t\sim 1$--3~My or the region around 0.7~AU at 0.1~My) the condition (\ref{condition}) is fulfilled.  

\begin{figure}[t!]
\centerline{\includegraphics[width=9.cm]{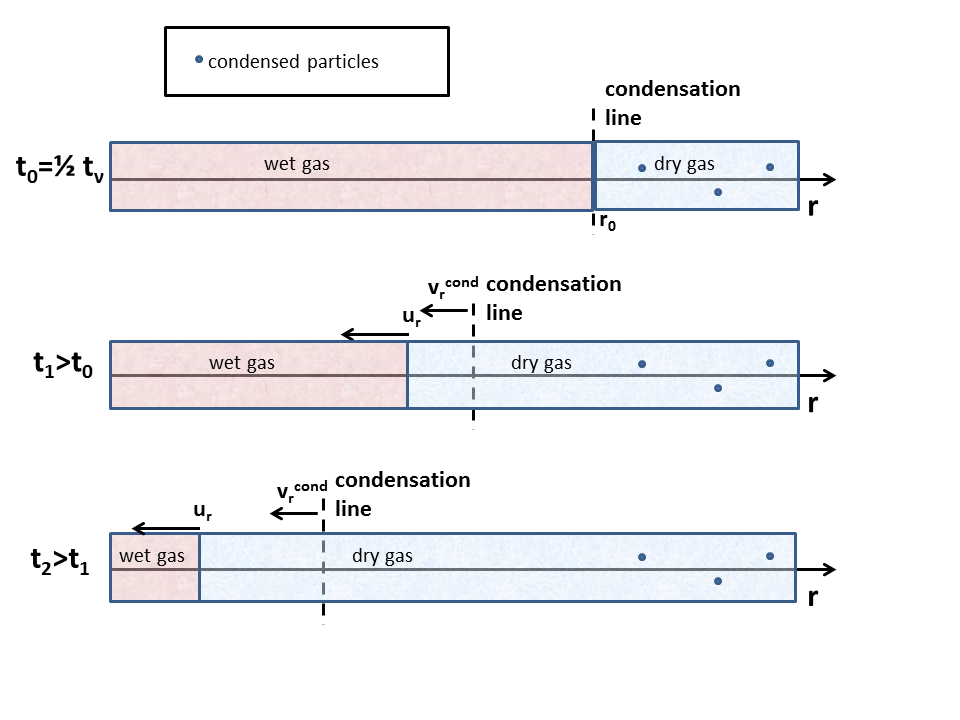}\includegraphics[width=9.cm]{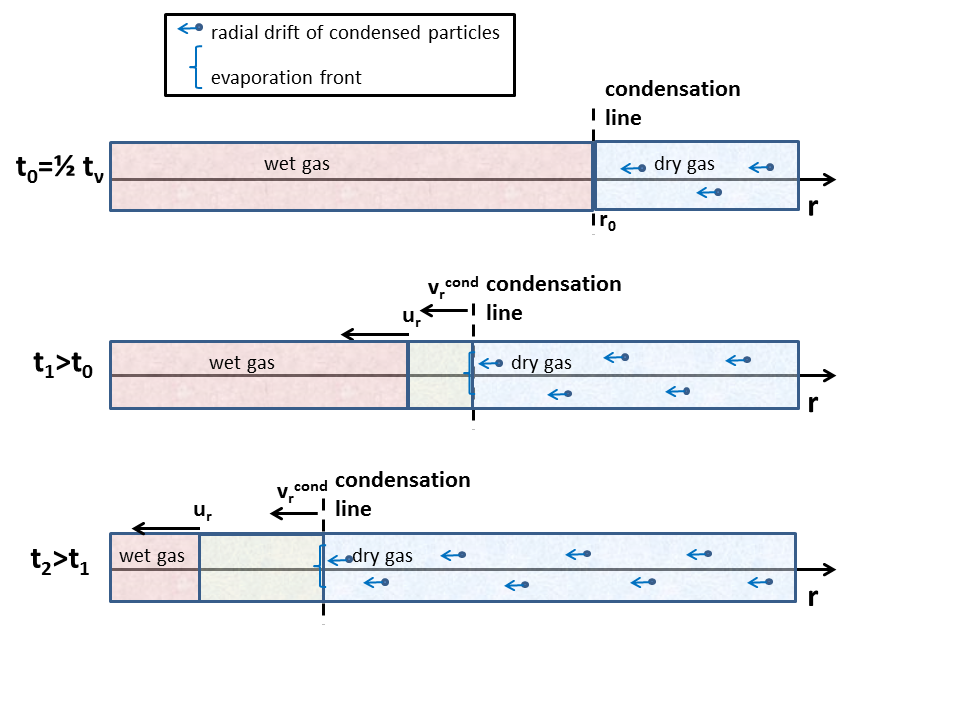}}
\caption{\footnotesize Sketch of the radial motion of a condensation line and of the gas, averaged over the vertical direction in the disk. {The left/right parts of the figure differ by the assumption that condensed particles do not drift/drift, respectively. From top to bottom,} each panel depicts the situation at different times, labeled on the left of each panel, with $t_0<t_1<t_2$. The time $t_0$ is defined as about half of the viscous timescale at the location of the condensation line $r_0$. The main horizontal arrow indicates the radial direction. The vertical dashed line shows the location of the condensation line as a function of time, approaching the star with a speed $v_r^{cond}$. The orange shaded region shows the ``wet'' gas, rich in the vapor of the considered species. The blue shaded region shows the ``dry'' gas, depleted in the vapor of considered species because the latter has condensed out. The outer boundary of the  wet gas region  also moves towards the star, with speed $u_r$. Because $u_r>v_r^{cond}$ this boundary moves away from the condensation line. {Thus, if there is no radial drift of particles (left part of the plot) the condensation line moves in a disk of dry gas, and therefore the amount of material that can condense out locally is negligible. Instead, in the case shown in the right part of the plot, the condensed particles repopulate, by radial drift, the disk down to the position of the condensation line. Also, the region (green) in between the ``wet'' and ``dry'' domains is resupplied in vapor of the considered species by particles sublimating when they pass through the condensation line (see the symbol for the sublimation front). Thus, the mass-flux of particles governs the abundance of the considered species in gaseous or solid form on both sides of the condensation line.} }
\label{snow-sketch}
\end{figure}

This result has an important implication. For $t\lesssim 1/2 t_\nu$ the condensation lines move very quickly. Thus there is the possibility that gas condenses out locally when the temperature drops. But for $t\gtrsim 1/2 t_\nu$ this process of direct condensation of gas loses importance.  This can be understood from the sketch in the {left part} of Fig.~\ref{snow-sketch}. Consider the location $r_0$ of a condensation line at time $t_0\sim 1/2 t_\nu$, where $t_\nu\equiv r_0^2/\nu$. The gas beyond the condensation line ($r>r_0$) is ``dry'', in the sense that the considered species is in condensed form; instead the gas at $r<r_0$ is ``wet'' in the sense that it is rich in the vapor of the considered species. {Now, consider first the idealized case where the condensed particles are large enough to avoid radial drift.} Because the radial drift of the gas is faster than the radial motion of the condensation line, the outer radial boundary of the  wet region  moves away from the condensation line, in the direction of the star. In reality the boundary between the two gases in wet and dry form is fuzzy, because it is smeared by turbulent diffusion. But it is clear, from the difference in radial velocities and a simple process of dilution, that the gas just inwards of the condensation lines has to become more and more depleted in the considered species as time progresses. Thus, as the condensation line advances towards the star, the amount of mass that can condense locally is very limited. {Thus, the condensed material can be (mostly) found only beyond the original location $r_0$ of the condensation line.}

Does this mean that a region of the disk originally too hot for a species to condense will remain depleted in that species forever, even if the temperature eventually drops well below the condensation threshold? In principle no, because {in a more realistic case (at least some of) the condensed particles are small enough to drift inwards by gas-drag, so that they can populate any region that has become cold enough to host them in solid form (see the right part of Fig.~\ref{snow-sketch}). Also, particles drifting through the condensation line can sublimate, thus resupplying the gas of the considered species in vapor form}. Thus, particle drift is the key to understanding the condensation line problem.

A solid particle can be characterized by a dimensionless parameter called the {\it  Stokes number}:
\begin{equation}
\tau_f={{\rho_b R}\over{\rho_g c_s}}\Omega\ ,
\label{stokes}
\end{equation}
where $\rho_b$ is the bulk density of the particle, $\rho_g$ is the density of the gas, $R$ is the size of the particle and $c_s$ is the sound speed. {A particle feels a wind from the gas, which has two components. The azimuthal component is due to the fact that the gas rotates around the star at a sub-Keplerian speed due to the pressure gradient in the disk; the radial component is due to the fact that the gas flows towards the Star, due to its own viscous evolution. Both components cause the radial drift of the particles towards the star at the speed (Weidenschilling, 1977; Takeuchi and Lin, 2002):}
\begin{equation}
v_r=-2 {{\tau_f}\over{\tau_f^2+1}}\eta v_K + {{u_r}\over{\tau_f^2+1}}\ ,
\label{vpart}
\end{equation}
where $v_K$ is the Keplerian velocity, {$u_r$ is the radial velocity of the gas} and $\eta$ is a measure of the gas pressure support:
\begin{equation}
\eta=-{{1}\over{2}}\left(H\over r\right)^2 {{{\rm d}\log P}\over{{\rm
      d}\log r}} \ .
\label{eta}
\end{equation}
Eq. (\ref{vpart}) is the final radial speed we looked for in this section.  {For mm-size particles or larger, typically the first term in (\ref{vpart}) dominates over the second one.}

The radial speed of particles is very fast. In fact, the typical value of $\eta$ is $\sim 3\times 10^{-3}$ so that the drift speed at 1~AU is $\sim 4 \tau_f \times 10^{-2}$~AU/y. Even particles as small as a millimeter ($\tau_f\sim 10^{-3}$ at $\sim 1$~AU) would travel most of the radial extent of the disk within the disk's lifetime. Solid particles condensed in the outer disk are therefore expected to potentially be delivered in the inner disk.

In conclusion, solving the snowline problem, i.e. understanding why Solar System objects remained depleted in species that should have condensed locally before the removal of the gas-disk, requires finding mechanisms that either prevent the radial drift of particles or inhibit the accretion of these particles onto pre-existing objects. Below we investigate some mechanisms, focussing on the cases of the snowline and the silicate line.

\section{The snowline}

In this section we discuss several mechanisms that could have potentially prevented the drift or the accretion of icy particles in the asteroid belt and the terrestrial planet zone even after that the snowline had passed across these regions. 

\subsection{Fast growth}

If icy particles had accreted each other quickly after their condensation, forming large objects (km-size or more) that were insensitive to gas drag, the inner Solar System would have received very little flux of icy material from the outer part of the disk (as in the example illustrated in the left part of Fig.~\ref{snow-sketch}).

We think that this scenario is unlikely. {The growth of planetesimals should have been extremely efficient for the fraction of the leftover icy particles to be small enough to have a negligible effect on the chemistry of the inner Solar System bodies. Such an efficient accretion has never been demonstrated in any model.} 

Observational constraints suggest the same, by showing that disks are dusty throughout their lifetime (see Williams and Cieza, 2011 or Testi et al., 2014 for reviews), {with the exception of the inner part of transitional disks (Espaillat et al., 2014) that we will address in sect.~4.4.} The Solar System offers its own constraints against this scenario. In chondrites, the ages of the individual chondrules inside the same meteorite span a few millions of years (Villeneuve et al., 2009; Connelly et al., 2012). Despite this variability, it is reasonable to assume that all particles (chondrules, CAIs,...) in the same rock got accreted at the same time. Thus, the spread in chondrule ages implies that particles were not trapped in planetesimals as soon as they formed; instead they circulated/survived in the disk for a long time before being incorporated into an object. Similarly, CAIs formed earlier than most chondrules (Connelly et al., 2012; Bollard et al., 2014), but they were incorporated in the meteorites with the chondrules; this means that the CAIs also spent significant time in the disk before being incorporated in macroscopic objects.  {Thus, it seems unlikely that virtually all icy particles had been accreted into planetesimals at early times, given that this did not happen for their refractory counterparts (CAIs and chondrules).}  

\subsection{Inefficient accretion}

A second possibility could be that the water-rich particles that drifted into the asteroid belt once the latter became cold enough,  were very small. Small particles accrete inefficiently on pre-existing planetesimals because they are too coupled with the gas (Lambrechts and Johansen, 2012; Johansen et al., 2015) and they are also very inefficient in triggering the streaming instability (Youdin and Goodman, 2005; Bai and Stone, 2010a,b; Carrera et al., 2015). Also, small particles are not collected in vortices, but rather accumulate in the low-vorticity regions at the dissipation scale of the turbulent cascade (Cuzzi et al., 2001). The levels of concentration that can be reached, however, are unlikely to be large enough to allow the formation of planetesimals (Pan et al., 2011). Thus, if the flux of icy material through the asteroid belt and the terrestrial planet region was mostly carried by very small particles, very little of this material would have been incorporated into asteroids and terrestrial planets precursors. But how small is small?

Again, chondrites give us important constraints. Chondrites are made of chondrules, which are 0.1--1 mm particles. Thus, particles this small could accrete into (or onto -- Johansen et al., 2015) planetesimals. The ice-rich particles flowing from the outer disk  are not expected to have been smaller than chondrules. Lambrechts and Johansen (2014) developed a model of accretion and radial drift of particles in the disk based on earlier work from Birnstiel et al. (2012). They found that the size of particles available in the disk decreased with time (the bigger particles being lost faster by radial drift). They estimated that, at 1 My, the particles at 2 AU were a couple of cm in size, so more than 10 times the chondrule size; the particles would have been chondrule-size at $\sim 10$~My. Thus, we don't see any reason why the asteroids should have accreted chondrules but not ice-rich particles, if the latter had drifted through the inner part of the asteroid belt. Consequently, this scenario seems implausible as well. 

\subsection{Filtering by planetesimals}

Particles, as they drifted radially, passed through a disk which presumably had already formed planetesimals of various sizes. Each planetesimal accreted a fraction of the drifting particles. If there were many planetesimals and they accreted drifting particles efficiently, the flow of icy material could have been decimated before reaching the inner Solar System region. 

Guillot et al. (2014) developed a very complete analytic model of the process of filtering of drifting dust and pebbles by planetesimals.  Unfortunately, the results are quite disappointing from our perspective. As shown by Figs. 21 and 22 of Guillot et al., in general only large boulders (about 10m in size) drifting from the outer disk (35~AU for the calculations illustrated in those figures, but the result is not very sensitive on this parameter) would have been accreted by planetesimals before coming within a few AU from the Sun. 

There are a few exceptions to this, however, also illustrated in the Guillot et al. paper. If the disk hosted a population of km-size planetesimals with a total mass corresponding to the solid mass in the Minimum Mass Solar Nebula model (MMSN; Weidenschillig, 1977b; Hayashi, 1981) and the turbulent stirring in the disk was weak ($\alpha=10^{-4}$ in the prescription of Shakura and Sunyaev, 1973) particles smaller than a millimeter in size could have been filtered efficiently and failed to reach the inner Solar System (however, see Fig. S2 in Johansen et al., 2015, for a different result). We think that it is unlikely that these parameters are pertinent for the real protoplanetary disk. In fact, we have seen above that icy particles are expected to have had sizes of a few cm at 1~My at 2~AU (Lambrechts and Johansen, 2014). Moreover, we believe it is unlikely that the size of the planetesimals that carried most of the mass of the disk was about 1~km. The formation of km-size planetesimals presents unsolved problems (e.g. the m-size barrier -Weidenschilling, 1977- and the bouncing barrier -G{\"u}ttler et al., 2009). Instead, modern accretion models (e.g. Johansen et al., 2007, 2015; Cuzzi et al., 2010) and the observed size distributions in the asteroid belt and the Kuiper belt suggest that planetesimals formed from self-gravitating clumps of small particles, with characteristics sizes of 100km or larger (Morbidelli et al., 2009).

Therefore, more interesting is the other extreme of the parameter space identified by Guillot et al. (2014). If a MMSN mass was carried by ``planetesimals'' more massive than Mars and the turbulent stirring was small (again, $\alpha=10^{-4}$), particles larger than 1~cm in size would have been filtered efficiently and would have failed to reach the inner Solar System. The lower limit of the mass of the filtering planetesimals decreases to 1/10 of a Lunar mass if the ``particles'' were meter-size boulders. Clearly, this is an important result. It is unclear, however, whether the protoplanetary disk could host so many planetesimals so big in size. The mass in solids in the MMSN model between 1 and 35 AU is about 50~$M_\oplus$. Assuming that this mass was carried by Mars-mass bodies would require the existence of about 500 of these objects.

\subsection{Filtering by proto-Jupiter}

At some point in the history of the protoplanetary disk, Jupiter started to form. The formation of the giant planets is not yet very clearly understood, so it is difficult to use models to assert when and where Jupiter had a given mass. 

However, it has been pointed out in Morbidelli and Nesvorny (2012) that when a planet reaches a mass of the order of 50 Earth masses it starts opening a partial gap in the disk. In an annulus just inside the outer edge of the gap the pressure gradient of the gas is reversed. Therefore, in this annulus the rotation of the gas around the Sun becomes faster than the Keplerian speed. Thus, the drag onto the particles is reversed. Particles do not spiral inwards, but instead spiral outwards. Consequently, particles drifting inwards from the outer disk have to stop near the outer edge of the gap. {This process is often considered to be at the origin of the so-called ``transitional disks'' (Espaillat et al., 2014), which show a strong depletion in mm-sized dust inside of some radius, with no proportional depletion in gas content.}

This mechanism for stopping the radial drift of solid particles has been revisited in Lambrechts et al. (2014), who used three dimensional hydro-dynamical simulations to improve the estimate of the planet's mass-threshold for reversing the gas pressure gradient. They found that the mass-threshold scales with the cube of the aspect ratio $h$ of the disk and is:
\begin{equation}
M_{iso}= 20M_\oplus \left({{h}\over{0.05}}\right)^3\ ,
\end{equation}
quite insensitive to viscosity (within realistic limits). Only particles very small and well-coupled with the gas (about 100$\mu$m or less; Paardekooper and Mellema, 2006) would pass through the gap opened by the planet and continue to drift through the inner part of the disk. {However, these particles are difficult to accrete by planetesimals, because they are ``blown in the wind'' (Guillot et al., 2014). Thus, they are not very important for the hydration of inner Solar System bodies. The particles which would be potentially important are those mm-sized or larger, because for these sizes the pebble accretion process is efficient (Ormel and Klahr 2010; Lambrechts and Johansen, 2012); however, for these particles the gap’s barrier is effective. }

According to Bitsch et al. (2015), when the snowline was at 3~AU (disk's accretion rate $\dot{M}\sim 3\times 10^{-8} M_\odot/y$) the aspect ratio of the disk was around 0.05 up to $\sim 10~AU$. So, the mass of 20~$M_\oplus$ is the minimum value required for the proto-Jupiter in order to stop the drift of icy pebbles and large grains. Basically, the constraint that asteroids and the precursors of the terrestrial planets did not accrete much ice  translates into a constraint on the mass and location of the proto-Jupiter. More specifically, the proto-Jupiter needs to have reached 20~$M_\oplus$ before the disk dropped below an accretion rate of  $\dot{M}\sim 3\times 10^{-8} M_\odot/y$ and it needs to have remained beyond the asteroid belt (i.e. beyond $\sim 3$~AU) until all the asteroids formed (about 3 My after CAIs). We think that this scenario is reasonable and realistic given that (i) Jupiter exists and thus it should have exceeded 20~$M_\oplus$ well within the lifetime of the disk and (ii) Jupiter is beyond 3~AU today. Nevertheless, there are important migration issues for Jupiter, that we will address in the Appendix. 

In this scenario, the current chemical structure of the Solar System would reflect the position of the snowline fossilized at the time when Jupiter achieved $\sim$20~$M_\oplus$\footnote{In reality, at this time there may still have been a reservoir of icy particles between the snowline location and the orbit of Jupiter. However, because of the typical drift rate of particles (10cm/s for mm-size dust, 1m/s for cm-size pebbles), if this reservoir was just a few AU wide (see Appendix), it should have emptied in $10^4$--$10^5$y, i.e. before that the snowline could move substantially.}. Therefore it makes sense that the fossilized snowline position corresponds to a disk already partially evolved (i.e. with $\dot{M}$ of a few $10^{-8} M_\odot/y$, instead of a few $10^{-7} M_\odot/y$, typical of an early disk), given that it may take considerable time (up to millions of years) to grow a planet of that mass. 

We also notice that this scenario is consistent, at least at the qualitative level, with the fact that ordinary and enstatite chondrites contain {\it some}  water (typically less than 1\% by mass, well below solar relative abundance of $\sim$ 50\%) and show secondary minerals indicative of water alteration (Baker et al., 2003; Alexander et al., 1989). In fact, it is conceivable that some particles managed to jump across the orbit of the proto-Jupiter (either because they were small enough to be entrained in the gas flow or by mutual scattering, once sufficiently piled up at the edge of Jupiter's gap). Because the entire asteroid region presumably had a temperature well below ice-sublimation by the time these chondrites formed, the icy particles that managed to pass through the planet's orbit would have been available in the asteroid belt region to be accreted. Of course, if most of the particles were retained beyond the orbit of Jupiter, the resulting abundance of water in these meteorites would have remained well below 50\% by mass, as observed. 

{The barrier to particle radial drift induced by the presence of the proto-Jupiter would not just have cut off the flow of icy material. It would also have cut off the flow of silicates. Thus, once the local particles had drifted away, the accretion of planetesimals should have stopped everywhere inside the orbit of the planet. Thus, it seems natural to expect that the chondrites should have stopped accreting at about the time of fossilization of the snowline position.  As we said in the introduction, chondrites accreted until $\sim 3$~My after CAI formation. Thus the formation of a 20~$M_\oplus$ proto-Jupiter should have occurred at about 3~My. Because the position of the snowline should have been at about 3~AU when this happened, this implies that the solar protoplanetary disk had an accretion rate $\dot{M}\sim 3\times 10^{-8} M_\odot/y$ (to sustain a snowline at 3~AU) at $\sim$3~My. Therefore, our disk was not typical (typical disks have a lower accretion rate at that age; Hartmann et al., 1998), although still within the distribution of observed accretion rates at this age (Manara et al., 2013).

There is, however, a second possibility. If there had been some mechanism recycling particles (i.e. sending the particles back out once they came close enough to the star), {or producing new particles in situ,} it is possible to envision that chondrules continued to form and chondrites continued to accrete them even after the flow of particles from the outer disk was cut off. In that case, the proto-Jupiter should still have formed when the snowline was at about 3 AU, but we would not have any chronological constraint on when this happened. It could have happened significantly earlier than 3 My. Thus, the solar protoplanetary disk might have had a typical accretion rate as a function of time.}

{A mechanism for producing small particles in situ is obviously collisions between planetesimals. Chondrules have been proposed to have formed as debris from collisions of differentiated planetesimals (Libourel and Krot, 2007; Asphaugh et al., 2011), although this is still debated (see e.g. Krot et al., 2009 for a reivew).}

Several mechanisms leading to a recycling of particles have been proposed, such as x-winds (Shu et al. 1996, 1997, 2001), gas outflow on the midplane (Takeuchi and Lin 2002; Ciesla 2007; Bai and Stone, 2010a) and disk winds  (Bai, 2014; Staff et al., 2014). Independently of the correct transport mechanism(s), the very detection of high temperature materials within comets (Brownlee et al. 2006; Nakamura et al. 2008; Bridges et al. 2012) demonstrates strong transport from the inner disk regions to the outer disk region. {However, it is not clear at which stage of the disk's life this outward transport was active and whether it concerned also particles larger than the microscopic ones recovered in the Stardust samples. }

{If a mechanism for recycling/producing particles in the inner disk really existed, another implication would be that planetesimals on either side of Jupiter's orbit eventually accreted from distinct reservoirs. The planetesimals inside of the orbit of the planet accreted only material recycled from the inner disk; instead the planetesimals outside of the proto-Jupiter's orbit accreted outer disk material, although possibly contaminated by some inner-disk material transported into the outer disk. In this respect, it may not be a coincidence that ordinary and carbonaceous chondrites appear to represent distinct chemical and isotopic reservoirs (Jacquet et al. 2012) because, in addition to the water content, these meteorites show two very distinct trends in the $\Delta^{17}$O-$\epsilon^{54}$Cr$^*$ isotope space (Warren et al. 2011). Today the parent bodies of both classes of meteorites reside in the asteroid belt, i.e. inside of the orbit of Jupiter. But in the {\it Grand Tack} scenario (Walsh et al., 2011) the parent bodies of the carbonaceous chondrites formed beyond Jupiter's orbit and got implanted into the asteroid belt during the phase of Jupiter's migration.}

\section{The silicate line}

According to terrestrial planet formation models, the small mass of Mercury and the absence of planets inside its orbit can be explained only by postulating that the disk of planetesimals and planetary embryos had an inner edge near 0.7~AU (Hansen, 2009). We argued in the introduction that this inner edge might reflect the location of the silicate condensation line at a very early age of the disk. But what could have prevented particles from drifting inside of 0.7~AU, once the disk there had cooled below the silicate condensation temperature? Obviously no giant planets formed near 0.7~AU, so the scenario invoked for the snowline cannot apply to this case.

A solution can be found in the results presented in Johansen et al., 2015. The authors pointed out that the very early disk is the most favorable environment for the production of planetesimal seeds via the streaming instability. This is because the streaming instability requires the presence of large particles, with Stokes numbers of the order of 0.1--1 {(e.g. significantly larger than chondrule-size particles)}. These particles drift very quickly in the disk, so they are rapidly lost. As shown in Lambrechts and Johansen (2014) the mass ratio between solid particles and gas, as well as the size of the dominant particles, decrease with time. Thus, the streaming instability becomes more and more unlikely to happen as time progresses. {Therefore, Johansen et al. argue that planetesimals formed in two stages. In the first stage planetesimal seeds formed by the streaming instability, triggered by large particles (decimeter across). This stage lasted for a short time only, due to the rapid loss of these large particles by radial drift. In the second stage the planetesimal seeds kept growing by accreting chondrule-sized particles, the only ones surviving and still drifting in the disk after a few My.}

{This model of planetesimal formation would provide a natural explanation for the fossilization of the silicate line.} Imagine that, when the phase of streaming instability was over, the accretion rate in the protoplanetary disk was $\dot{M}\sim 1.5\times 10^{-7} M_\odot/y$. In this case, the silicate sublimation line was at $\sim 0.7$~AU. Thus, presumably no planetesimal seeds could have formed within this radius, because until that time only the very refractory material would have been available in solid form there (a small fraction of the total mass, insufficient to trigger the streaming instability). After the streaming instability phase was over, planetesimals continued to grow by the accretion of smaller particles onto the planetesimal seeds (Johansen et al., 2015). Meanwhile the disk cooled so that silicate particles could drift within 0.7~AU and remain solid. However, because of the absence of planetesimal seeds within this radius, they could not be accreted by anything, and thus they simply drifted towards the Sun.  Clearly, in this scenario the final planetesimal disk would have an inner edge at 0.7~AU, fossilizing the early location of the silicate line that characterized the formation of the planetesimal seeds. 

Admittedly, this scenario is still qualitative, but it is seducing in its simplicity. The possibility to explain location of the inner edge of the planetesimal disk by this mechanism is an additional argument in favor of the two-phases model for planetesimal growth, simulated in Johansen et al. (2015).  

\section{Conclusions}

The chemical composition of the objects of the Solar System seems to reflect a condensation sequence set by a temperature gradient typical of an ``early'' disk, still significantly warm in its inner part. Particularly significant is the situation concerning water. The terrestrial planets and the asteroids predominant in the inner belt are water-poor. However, the disk's temperature should have decreased below the water-condensation level even at 1 AU before the disappearance of the gas. So, the question why terrestrial planets and asteroids are not all water rich is a crucial one (Oka et al., 2011). Interestingly, water is not the only chemical species that reveals this conundrum. Inner Solar System objects are in general more depleted in volatile element than they should be, given the temperatures expected in the disk at its late stages. Also, terrestrial planet formation models argue that, in order to explain the small mass of Mercury, the planetesimal disk had to have an inner edge at about 0.7~AU (Hansen, 2009; Walsh et al., 2011). This boundary could correspond to the location of the evaporation front for silicates, but again only for a massive (i.e. ``early'') disk.

In this Note we have discussed how the composition of Solar System objects could reflect the location of the condensation lines fossilized at some specific stages of the disk's evolution. Some mechanisms have been quickly dismissed, others look promising and we propose them to the community for further discussion and investigation. 

\begin{figure}
\centerline{\includegraphics[height=14.cm]{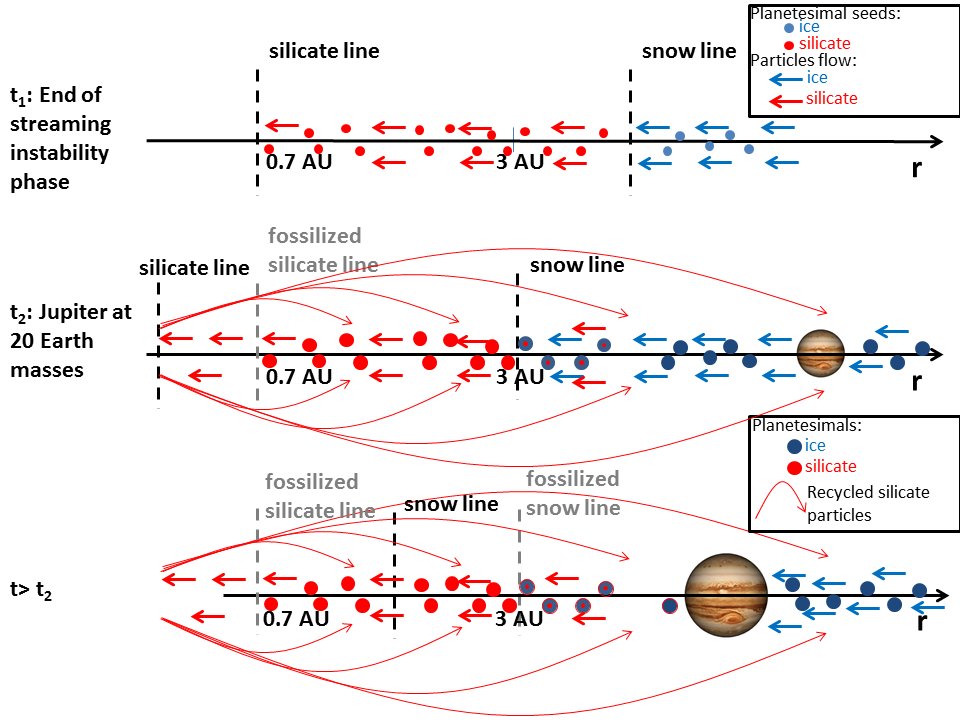}}
\caption{\small Sketch of the solution to the follisized condensation lines problem proposed in this paper. The top and central panels show the situation at the times when the silicate line, first, then the snowline, remain fossilized. The bottom panel sketches the situation after the fossilization of the snowline, under the assumption where accretion of planetesimals in the inner disk continues thanks to the recycling of small particles in outwards flows. {If a recycling or particle-generation mechanism did not exist, all planetesimals inside of the orbit of proto-Jupiter should have stopped accreting at the time of fossilization of the snowline.} See text for detailed description.}
\label{sketch-final}
\end{figure}

First, we have demonstrated that the radial motion of gas towards the central star is faster than the inward motion of the condensation lines. This implies that there cannot be condensation of gas in a region swept by the motion of a condensation line, even if this may sound paradoxical. This is because the gas in the considered region comes from farther out and therefore it should have already condensed there (see Fig.~\ref{snow-sketch}, left side). Thus, the enrichment of a disk region in condensed elements when the temperature drops can only be due to the radial drift of solid particles from the more distant disk (see Fig.~\ref{snow-sketch}, right side).

With this consideration in mind, the scenario that we propose to explain the fossilized condensation lines is the one sketched in Fig.~\ref{sketch-final}. Planetesimal formation occurred in two stages, as proposed in Johansen et al. (2015). In the very early disk, dust coagulation produced pebbles and boulders of sizes ranging from decimeters to, possibly, a meter. These objects were very effective in triggering the streaming instability (Youdin and Goodman, 2005; Johansen et al., 2007; Bai and Stone, 2010a,b) and areodynamically clumped together forming planetesimal seeds of about 100~km in size.  This phase, however, could not last long, because these boulders drifted quickly through the disk and those that were not rapidly incorporated into a planetesimal seed got lost by drifting into the Sun. We propose that, when this stage ended, the silicate condensation line was at 0.7~AU. This would correspond to a disk with an accretion rate of $\dot{M}\sim 1.5\times 10^{-7} M_\odot/y$, for a nominal metallicity of 1\%. Then, no planetesimal seeds could have formed within this radius, because of the lack of a sufficient amount of solids. In the second stage, the surviving solid particles were too small and their mass ratio with the gas too low to trigger streaming instability (Lambrechts and Johansen, 2014). Thus, these particles could only be accreted onto the already formed planetesimal seeds (Johansen et al., 2015). If no seeds existed inside of 0.7~AU, this radius remained the inner edge of the planetesimal disk, a fossil trace of the silicate line at the end of the first stage of planetesimal growth. 

Meanwhile the temperature in the disk continued to decrease. The snowline moved towards the Sun. The region swept by the snowline became increasingly enriched in icy material due to the radial drift of particles from the outer disk. When the mass of the proto-Jupiter reached 20~$M_\oplus$, however, this flux of icy particles stopped. The opening of a shallow gap by the proto-planet created a barrier to the inward drift of the particles by gas drag. Thus, the flux of icy material across Jupiter's orbit was interrupted, {presumably making our Solar System look like a ``transitional disk''.} We propose that the proto-Jupiter reached this critical mass when the snowline was at about 3~AU. This corresponds to a disk with a stellar accretion rate of $\dot{M}\sim 3\times 10^{-8} M_\odot/y$, assuming the canonical metallicity of 1\%. Thus, the objects inside 3~AU, which could not accrete ice up to that time because the temperature was too high, could not accrete a significant amount of ice also after that the temperature dropped, because of the interrupted icy-particle flow. Instead, they could have continued to accrete refractory particles if the latter were recycled {or continuously reproduced} in the disk, thanks to the existence of outwards flows in the midplane (Takeuchi and Lin, 2002; Ciesla, 2007; Bai and Stone, 2010a), x-winds (Shu et al., 1996,1997,2001), disk winds (Bai, 2014; Staff et al., 2014) {or collisions (Libourel and Krot, 2007; Asphaugh et al., 2011)}. Thus, the resulting chemistry of planetesimals reflect the location of the snowline fossilized at the time the proto-Jupiter reached 20~$M_\oplus$. 

Notice that, because it takes more time to form the proto-Jupiter than the planetesimal seeds, it makes sense that the snowline appears fossilized at the location corresponding to a ``later'' disk than the silicate line ($\dot{M}\sim 3\times 10^{-8} M_\odot/y$ instead of $\dot{M}\sim 1.5\times 10^{-7} M_\odot/y$). 

This scenario is much simpler than what has been proposed so far for the snowline problem or the origin of the inner edge of the planetesiamal disk (reviewed in Sect.~2). It is indeed appealing for its simplicity. 

This scenario leads to a few predictions. For the Solar System it predicts that the condensation lines corresponding to species much more volatile than water (e.g. CO, with a condensation temperature of $\sim 25$~K) should not have been fossilized because only Jupiter and Saturn are sufficiently massive to stop the flow of drifting particles and these planets should always have been too close to the Sun. Thus, the composition of outer Solar System bodies should reflect the location of these lines at the end of the disk's lifetime. Perhaps this can explain the compositions of Uranus and Neptune (Ali-Dib et al., 2014).

For extrasolar planetary systems, the scenario predicts that systems without giant planets should be much more volatile rich in their inner parts than a system like ours. This seems consistent with the observations, which show a large number of systems of low-density super-Earths in close-in orbits and no giant planets farther out (Fressin et al., 2013). In principle the low bulk-densities could be explained by the presence of extended H and He atmosphere around rocky planets. But Lopez and Fortney (2014) concluded that the observed size distribution of extrasolar planets, with a sharp drop-off above 3 R$_\oplus$, is diagnostic that most super-Earths are water-rich. In fact, refractory planets with extended atmospheres would have a more uniform size distribution.

{Finally, the fossilization of the silicate line should be a generic process, although the location at which this condensation line is fossilized may change from disk to disk depending on the duration of the streaming instability stage and the evolution of the temperature in that timeframe. This suggests that ``hot'' extrasolar planets (with orbital radii significantly smaller than Mercury's) did not form in situ but migrated to their current orbits from some distance away.}

Clearly, the scenario proposed in this Note remains speculative. However, 
with the improved understanding of disk evolution and its chronology, planetesimal accretion and giant planet growth, it will be possible in a hopefully not distant future to test it on more quantitative grounds against the available constraints.

\section{Appendix: Planet migration issues}
\label{migration}

Planets are known to migrate in disks (see Baruteau et al., 2014 for a review).  Thus, we discuss here possible scenarios that could explain how the proto-Jupiter remained beyond $\sim 3$~AU until the chondrite formation time. 

A first possibility is that Jupiter's core started to form sufficiently far in the disk, so that it could not reach 3~AU before  $\sim 3$~My. This is one of the approaches taken in Bitsch et al., 2015b. In their model, Jupiter started growing by pebble accretion at about 20~AU. 

A second possibility is offered by the subtle action of the entropy-driven corotation torque (Paardekooper and Mellema, 2006b; Paardekooper et al., 2010, 2011; Masset and Casoli, 2009, 2010). This torque can reverse the migration of intermediate-mass planets (several Earth masses) in localized regions of the disk where the temperature gradient is steep. Bitsch et al. (2014a, 2015) showed migration maps as a function of location and planet mass at different evolutionary stages (i.e. different values of $\dot{M}$) of the disk.  They found that the outward migration region is adjacent to the snowline. It typically ranges from the snowline location up to a few AUs beyond the snowline. For an early disk ($\dot{M}=7\times 10^{-8} M_\odot/y$) outward migration concerns planets with masses between 5 and 40 Earth masses. A proto-Jupiter in this mass range would therefore be retained at 6--8 AU (depending on its mass), the snowline being at $\sim 4$ AU (see Fig. 7 of Bitsch et al., 2015).  When the disk loses mass and cools, the outward migration region shifts towards the Sun with the snowline position. But it also shrinks in the planet-mass parameter space. In a late disk with $\dot{M}=8.75\times 10^{-9} M_\odot/y$ (snowline at $\sim 1 AU$) the outward migration region still extends to $\sim 3 AU$, but only for planet masses smaller than 10~$M_\oplus$. Thus, in principle, a pebble-stopping planet of 20~$M_\oplus$ should have been released to free Type-I migration and should have penetrated into the inner Solar System, possibly too early relative to the chondrite formation time of $\sim 3$~My. 

\begin{figure}[t!]
\centerline{\includegraphics[height=10.cm]{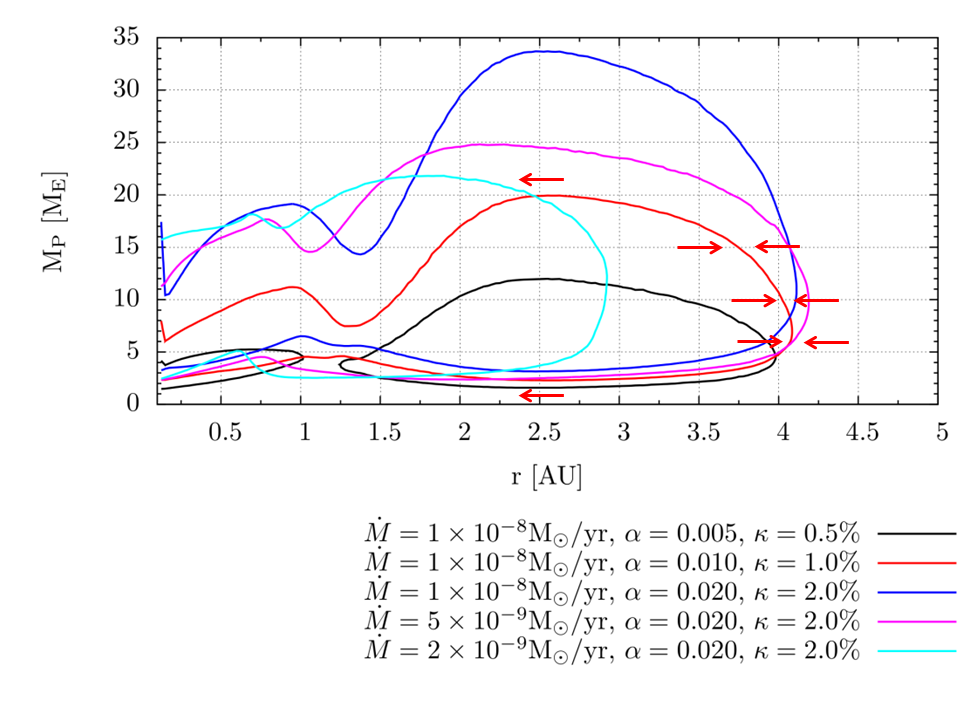}}
\caption{\small Contours of the outward migration region in the parameter space heliocentric distance vs. planetary mass. Each color corresponds to a different disk, whose parameters $\dot{M}, \alpha$ and $\kappa$ are reported on the plot. To help reading this plot, the red arrows show the direction of migration of planets of different masses and locations for the case with $\dot{M}=10^{-8} M_\odot$/y, $\alpha=0.01$ and $\kappa=$1\%, corresponding to the red contour. A planet of an appropriate mass (between 2.5 and 20 $M_E$ for the case of the red contour) migrating inwards from the outer disk would stop at the right-hand-side boundary of the outward migration region. The black and red contorus are too limited in planet-mass range to be able to trap a 20~$M_\oplus$ planet, but the other disks could retain a proto-Jupiter of this mass beyond 2.5 or 3 AU. }
\label{eggs}
\end{figure}

We stress, however, that the exact location and shape of the outward migration region depends on the adopted parameters for the disk, as shown in Fig.~\ref{eggs}. In particular,  the upper limit in planet mass for outward migration is due to torque saturation. It can be increased if the disk is more viscous. The size of the coorbital region of a planet (the one characterized by horseshoe streamlines) has a width $x_s\propto\sqrt{M_p}$, where $M_p$ is the planet mass. The timescale for viscous transport across this region is therefore 
\begin{equation}
t_\nu\propto x_s^2/\nu \propto M_p/\nu \ .
\label{tnu}
\end{equation}
The libration timescale in the horseshoe region is 
\begin{equation}
t_{lib}\propto {1\over\sqrt{M_p}}\ .
\label{tlib}
\end{equation} 
Saturation is achieved when $t_{lib}<<t_\nu$. 
If we want saturation to occur for a planet $N$ times bigger in mass, the scaling of (\ref{tnu}) and (\ref{tlib}) implies that $\nu$ has to be ${N}^{3/2}$ times bigger. 

However, to have the outward migration region in the same radial range we need that the thermal structure of the disk does not change with the increase in viscosity. Because the viscous transport in the disk is proportional to $\nu\Sigma$, we need that $\Sigma$ is ${N}^{3/2}$ smaller. And because the temperature in the disk is proportional to $(\kappa\nu\Sigma^2)^{1/4}$ (see \ref{temp}) we need that the opacity $\kappa$ (basically the mass ratio between micron-sized dust and gas) scales as ${N}^{3/2}$. In other words, a planet of 20~$M_\oplus$ can be retained at 3~AU in a disk with $\dot{M}=8.75\times 10^{-9} M_\odot/y$ if the viscosity is $\sim 3$ times higher and the opacity $\sim 3$ times larger than assumed in Bitsch et al. (2015).  Given the uncertainties on disk parameters, we cannot exclude this possibility.  Fig.~\ref{eggs} indeed shows ``late disks'' (i.e. with a small $\dot{M}$ -- a few $10^{-9} M_{\odot}/y$) with an outward migration region capable of retaining a 20~$M_\oplus$ planet beyond 2.5--3~AU.

Nevertheless, however we play with disk parameters, it is clear that a planet is eventually released to inward migration once it becomes massive enough. Thus, Jupiter should have eventually invaded the asteroid belt, (unless it started so far out in the disk that it was not able to reach the asteroid belt within the disk's lifetime; see some simulations in Bitsch et al., 2015b). The migration of Jupiter through the belt is contemplated in the so-called {\it Grand Tack scenario} (Walsh et al., 2011), in which Jupiter reached 1.5~AU before being pulled back to its current distance by the presence of Saturn. This scenario of inward-then-outward migration of Jupiter can explain the excitation and depletion of the asteroid belt and the abortion of the growth of Mars. Again, because chondritic planetesimals accrete until $\sim 3$~My, what is important is that Jupiter did not invade the asteroid belt till that time. 


\acknowledgments 

We acknowledge support by the French ANR, project number ANR-13--13-BS05-0003-01  projet MOJO (Modeling the Origin of JOvian planets). A.J. is grateful for the financial support from the European Research Council
(ERC Starting Grant 278675-PEBBLE2PLANET) and the Swedish Research Council (grant 2014-5775). A.J. and B.B. thanks the Knut and Alice Wallenberg Foundation for their financial support. S. J. was supported by the European Research Council (ERC) Advanced Grant ACCRETE (contract number 290568). We are grateful to C. Dullemond and an anonymous reviewer for their constructive reviews.

\section{References}

\begin{itemize}

\item[--] Albar{\`e}de, F.\ 2009.\ 
Volatile accretion history of the terrestrial planets and dynamic 
implications.\ Nature 461, 1227-1233.

\item[--] Alexander, C.~M.~O., 
Barber, D.~J., Hutchison, R.\ 1989.\ The microstructure of Semarkona and 
Bishunpur.\ Geochimica et Cosmochimica Acta 53, 3045-3057. 

\item[--] Alexander, R., 
Pascucci, I., Andrews, S., Armitage, P., Cieza, L.\ 2014.\ The Dispersal of 
Protoplanetary Disks.\ Protostars and Planets VI 475-496. 

\item[--] Ali-Dib, M., Mousis, 
O., Petit, J.-M., Lunine, J.~I.\ 2014.\ The Measured Compositions of Uranus 
and Neptune from their Formation on the CO Ice Line.\ The Astrophysical 
Journal 793, 9. 

\item[--] Asphaug, E., Jutzi, M., Movshovitz, N.\ 2011.\ Chondrule formation during planetesimal accretion.\ Earth and Planetary Science Letters 308, 369-379. 

\item[--] Bai, X.-N., Stone, 
J.~M.\ 2010a.\ Dynamics of Solids in the Midplane of Protoplanetary Disks: 
Implications for Planetesimal Formation.\ The Astrophysical Journal 722, 
1437-1459. 

\item[--] Bai, X.-N., Stone, 
J.~M.\ 2010b.\ The Effect of the Radial Pressure Gradient in Protoplanetary 
Disks on Planetesimal Formation.\ The Astrophysical Journal 722, L220-L223. 

\item[--] Bai, X.-N.\ 2014.\ 
Hall-effect-Controlled Gas Dynamics in Protoplanetary Disks. I. Wind 
Solutions at the Inner Disk.\ The Astrophysical Journal 791, 137.

\item[--] Bailli{\'e}, K., 
Charnoz, S., Pantin, {\'E}.\ 2015.\ Time evolution of snow regions and 
planet traps in an evolving protoplanetary disk.\ ArXiv e-prints 
arXiv:1503.03352. 

\item[--] Baker, L., Franchi, 
I.~A., Wright, I.~P., Pillinger, C.~T.\ 2003.\ Aqueous Alteration on 
Ordinary Chondrite Parent Bodies- The Oxygen Isotopic Composition of 
Water.O.\ EGS - AGU - EUG Joint Assembly 11198. 

\item[--]  Baruteau, C., Crida, 
A., Paardekooper, S.-J., Masset, F., Guilet, J., Bitsch, B., Nelson, R., 
Kley, W., Papaloizou, J.\ 2014.\ Planet-Disk Interactions and Early 
Evolution of Planetary Systems.\ Protostars and Planets VI 667-689.

\item[--] Batygin, K., 
Laughlin, G.\ 2015.\ Jupiter's Decisive Role in the Inner Solar System's 
Early Evolution.\ ArXiv e-prints arXiv:1503.06945. 


\item[--] Birnstiel, T., Klahr, H., Ercolano, B.\ 2012.\ A simple model for the evolution of the dust population in protoplanetary disks.\ Astronomy and Astrophysics 539, AA148. 

\item[--] Bitsch, B., Crida, A., Morbidelli, A., Kley, W., Dobbs-Dixon, I.\ 2013.\ Stellar irradiated discs and implications on migration of embedded planets. I. Equilibrium discs.\ Astronomy and Astrophysics 549, A124. 

\item[--] Bitsch, B., Morbidelli, A., Lega, E., Crida, A.\ 2014a.\ Stellar irradiated discs and implications on migration of embedded planets. II. Accreting-discs.\ Astronomy and Astrophysics 564, AA135. 

\item[--] Bitsch, B., Morbidelli, A., Lega, E., Kretke, K., Crida, A.\ 2014b.\ Stellar irradiated discs and implications on migration of embedded planets. III. Viscosity transitions.\ Astronomy and Astrophysics 570, AA75. 

\item[--] Bitsch, B., Johansen, A., Lambrechts, M., Morbidelli, A.\ 2015.\ The structure of protoplanetary discs around evolving young stars.\ Astronomy and Astrophysics 575, AA28. 

\item[--] Bitsch, B., Lambrechts, M., Johansen, A.,\ 2015b.\ The growth of planets by pebble accretion in evolving protoplanetary discs. Astronomy and Astrophysics, in press.

\item[--] Bollard, J., Connelly, 
J.~N., Bizzarro, M.\ 2014.\ The Absolute Chronology of the Early Solar 
System Revisited.\ LPI Contributions 1800, 5234. 

\item[--] Bridges, J.~C., Changela, H.~G., Nayakshin, S., Starkey, N.~A., Franchi, I.~A.\ 2012.\ Chondrule fragments from Comet Wild2: Evidence for high temperature processing in the outer Solar System.\ Earth and Planetary Science Letters 341, 186-194.

\item[--] Brownlee, D., and 182 
colleagues 2006.\ Comet 81P/Wild 2 Under a Microscope.\ Science 314, 1711. 

\item[--] Carrera, D., Johansen, 
A., Davies, M.~B.\ 2015.\ How to form planetesimals from mm-sized 
chondrules and chondrule aggregates.\ ArXiv e-prints arXiv:1501.05314. 

\item[--] Ciesla, F.~J.\ 2007.\ Outward 
Transport of High-Temperature Materials Around the Midplane of the Solar 
Nebula.\ Science 318, 613. 

\item[--] Chiang, E.~I., 
Goldreich, P.\ 1997.\ Spectral Energy Distributions of T Tauri Stars with 
Passive Circumstellar Disks.\ The Astrophysical Journal 490, 368-376. 

\item[--] Connelly, J.~N., 
Bizzarro, M., Krot, A.~N., Nordlund, {\AA}., Wielandt, D., Ivanova, M.~A.\ 
2012.\ The Absolute Chronology and Thermal Processing of Solids in the 
Solar Protoplanetary Disk.\ Science 338, 651. 

\item[--] Cossou, C., Raymond, S.~N., Hersant, F., Pierens, A.\ 2014.\ Hot super-Earths and giant planet cores from different migration histories.\ Astronomy and Astrophysics 569, AA56. 

\item[--] Cuzzi, J.~N., Hogan, 
R.~C., Paque, J.~M., Dobrovolskis, A.~R.\ 2001.\ Size-selective 
Concentration of Chondrules and Other Small Particles in Protoplanetary 
Nebula Turbulence.\ The Astrophysical Journal 546, 496-508. 

\item[--] Cuzzi, J.~N., Hogan, 
R.~C., Bottke, W.~F.\ 2010.\ Towards initial mass functions for asteroids 
and Kuiper Belt Objects.\ Icarus 208, 518-538. 

\item[--] Davis, S.~S.\ 2005.\ 
Condensation Front Migration in a Protoplanetary Nebula.\ The Astrophysical 
Journal 620, 994-1001. 

\item[--] Dullemond, C.~P., 
Dominik, C., Natta, A.\ 2001.\ Passive Irradiated Circumstellar Disks with 
an Inner Hole.\ The Astrophysical Journal 560, 957-969. 

\item[--] Dullemond, C.~P., van Zadelhoff, G.~J., Natta, A.\ 2002.\ Vertical structure models of T Tauri and Herbig Ae/Be disks.\ Astronomy and Astrophysics 389, 464-474. 

\item[--] Dullemond, C.~P.\ 2002.\ The 2-D structure of dusty disks around Herbig Ae/Be stars. I. Models with grey opacities.\ Astronomy and Astrophysics 395, 853-862. 

\item[--] Espaillat, C., 
Muzerolle, J., Najita, J., Andrews, S., Zhu, Z., Calvet, N., Kraus, S., 
Hashimoto, J., Kraus, A., D'Alessio, P.\ 2014.\ An Observational 
Perspective of Transitional Disks.\ Protostars and Planets VI 497-520. 

\item[--] Fressin, F., Torres, 
G., Charbonneau, D., Bryson, S.~T., Christiansen, J., Dressing, C.~D., 
Jenkins, J.~M., Walkowicz, L.~M., Batalha, N.~M.\ 2013.\ The False Positive 
Rate of Kepler and the Occurrence of Planets.\ The Astrophysical Journal 
766, 81. 

\item[--] Garaud, P., Lin, 
D.~N.~C.\ 2007.\ The Effect of Internal Dissipation and Surface Irradiation 
on the Structure of Disks and the Location of the Snow Line around Sun-like 
Stars.\ The Astrophysical Journal 654, 606-624. 

\item[--] Guillot, T., Ida, S., Ormel, C.~W.\ 2014.\ On the filtering and processing of dust by planetesimals. I. Derivation of collision probabilities for non-drifting planetesimals.\ Astronomy and Astrophysics 572, AA72. 

\item[--] G{\"u}ttler, C., 
Blum, J., Zsom, A., Ormel, C.~W., Dullemond, C.~P.\ 2009.\ The first phase 
of protoplanetary dust growth: The bouncing barrier.\ Geochimica et 
Cosmochimica Acta Supplement 73, 482. 


\item[--] Hansen, B.~M.~S.\ 2009.\ 
Formation of the Terrestrial Planets from a Narrow Annulus.\ The 
Astrophysical Journal 703, 1131-1140. 

\item[--] Hartmann, L., Calvet, 
N., Gullbring, E., D'Alessio, P.\ 1998.\ Accretion and the Evolution of T 
Tauri Disks.\ The Astrophysical Journal 495, 385-400. 

\item[--] Hayashi, C.\ 1981. Structure of the Solar
Nebula, Growth and Decay of Magnetic Fields and Effects of Magnetic
and Turbulent Viscosities on the Nebula.\ {\it Progress of Theoretical
Physics Supplement} 70, 35-53.

\item[--] Hubbard, A., Ebel, 
D.~S.\ 2014.\ Protoplanetary dust porosity and FU Orionis outbursts: 
Solving the mystery of Earth's missing volatiles.\ Icarus 237, 
84-96. 

\item[--] Hueso, R., Guillot, T.\ 2005.\ Evolution of protoplanetary disks: constraints from DM Tauri and GM Aurigae.\ Astronomy and Astrophysics 442, 703-725. 

\item[--] Ida, S., Lin, D.~N.~C.\ 
2008.\ Toward a Deterministic Model of Planetary Formation. IV. Effects of 
Type I Migration.\ The Astrophysical Journal 673, 487-501.

\item[--] Isella, A., Natta, A.\ 2005.\ The shape of the inner rim in proto-planetary disks.\ Astronomy and Astrophysics 438, 899-907. 

\item[--] Jacobson, 
S.~A., Morbidelli, A.\ 2014.\ Lunar and terrestrial planet formation in the 
Grand Tack scenario.\ Royal Society of London Philosophical Transactions 
Series A 372, 0174. 

\item[--]  Jacquet, E., Gounelle, 
M., Fromang, S.\ 2012.\ On the aerodynamic redistribution of chondrite 
components in protoplanetary disks.\ Icarus 220, 162-173.

\item[--] Johansen, A., Oishi, 
J.~S., Mac Low, M.-M., Klahr, H., Henning, T., Youdin, A.\ 2007.\ Rapid 
planetesimal formation in turbulent circumstellar disks.\ Nature 448, 
1022-1025. 

\item[--] Johansen, A., Mac Low, 
M.-M., Lacerda, P., Bizzarro, M.\ 2015.\ Growth of asteroids, planetary 
embryos and Kuiper belt objects by chondrule accretion. Science Advances: e1500109 

\item[--] Koenigl, A.\ 1991.\ Disk 
accretion onto magnetic T Tauri stars.\ The Astrophysical Journal 370, 
L39-L43. 

\item[--] Krot, A.~N., and 15 
colleagues 2009.\ Origin and chronology of chondritic components: A 
review.\ Geochimica et Cosmochimica Acta 73, 4963-4997. 

\item[--] Lambrechts, M., Johansen, A.\ 2014.\ Forming the cores of giant planets from the radial pebble flux in protoplanetary discs.\ Astronomy and Astrophysics 572, AA107. 

\item[--] Lambrechts, M., Johansen, A., Morbidelli, A.\ 2014.\ Separating gas-giant and ice-giant planets by halting pebble accretion.\ Astronomy and Astrophysics 572, AA35. 

\item[--] Libourel, G., Krot, A.~N.\ 2007.\ Evidence for the presence of planetesimal material among the precursors of magnesian chondrules of nebular origin.\ Earth and Planetary Science Letters 254, 1-8. 

\item[--] Lin, D.~N.~C., Bodenheimer, 
P., Richardson, D.~C.\ 1996.\ Orbital migration of the planetary companion 
of 51 Pegasi to its present location.\ Nature 380, 606-607. 

\item[--] Lodders, K.\ 2003.\ Solar 
System Abundances and Condensation Temperatures of the Elements.\ The 
Astrophysical Journal 591, 1220-1247. 

\item[--] Luu, T.H., Young, E.D., Gounelle, M., Chaussidon, M., 2015. Short time interval for condensation of high-temperature silicates in the solar accretion disk. PNAS 112, 1298-1303.

\item[--] Lynden-Bell, 
D., Pringle, J.~E.\ 1974.\ The evolution of viscous discs and the origin of 
the nebular variables.\ Monthly Notices of the Royal Astronomical Society 
168, 603-637. 

\item[--] Manara, C.~F., and 11 colleagues 2013.\ X-shooter spectroscopy of young stellar objects. II. Impact of chromospheric emission on accretion rate estimates.\ Astronomy and Astrophysics 551, AA107. 

\item[--] Martin, R.~G., Livio, 
M.\ 2012.\ On the evolution of the snow line in protoplanetary discs.\ 
Monthly Notices of the Royal Astronomical Society 425, L6-L9. 

\item[--] Martin, R.~G., Livio, 
M.\ 2013.\ On the evolution of the snow line in protoplanetary discs - II. 
Analytic approximations.\ Monthly Notices of the Royal Astronomical Society 
434, 633-638. 

\item[--] Marty, B.\ 2012.\ The origins and concentrations of water, carbon, nitrogen and noble gases on Earth.\ Earth and Planetary Science Letters 313, 56-66. 

\item[--] Masset, F.~S., 
Morbidelli, A., Crida, A., Ferreira, J.\ 2006.\ Disk Surface Density 
Transitions as Protoplanet Traps.\ The Astrophysical Journal 642, 478-487. 

\item[--] Masset, F.~S., 
Casoli, J.\ 2009.\ On the Horseshoe Drag of a Low-Mass Planet. II. 
Migration in Adiabatic Disks.\ The Astrophysical Journal 703, 857-876. 

\item[--] Masset, F.~S., 
Casoli, J.\ 2010.\ Saturated Torque Formula for Planetary Migration in 
Viscous Disks with Thermal Diffusion: Recipe for Protoplanet Population 
Synthesis.\ The Astrophysical Journal 723, 1393-1417. 

\item[--] McCubbin, F. M., Hauri, E. H., Elardo, S. M., Vander Kaaden, K. E., Wang, J., and Shearer, C. K. 2012. Hydrous melting of the martian mantle produced both depleted and enriched shergottites. Geology 40, 683-686.

\item[--] McDonough, W.F., Sun, S.S, 1995. The composition of the Earth. Chemical Geology 120, 223-253

\item[--] Morbidelli, A., Chambers, J., Lunine, J.~I., Petit, J.~M., Robert, F., Valsecchi, G.~B., Cyr, K.~E.\ 2000.\ Source regions and time scales for the delivery of water to Earth.\ Meteoritics and Planetary Science 35, 1309-1320. 

\item[--] Morbidelli, A., 
Bottke, W.~F., Nesvorn{\'y}, D., Levison, H.~F.\ 2009.\ Asteroids were born 
big.\ Icarus 204, 558-573. 

\item[--] Morbidelli, A., Nesvorny, D.\ 2012.\ Dynamics of pebbles in the vicinity of a growing planetary embryo: hydro-dynamical simulations.\ Astronomy and Astrophysics 546, AA18. 

\item[--]  Nakamura, T., and 11 
colleagues 2008.\ Chondrulelike Objects in Short-Period Comet 81P/Wild 2.\ 
Science 321, 1664. 

\item[--] O'Brien, D.~P., 
Morbidelli, A., Levison, H.~F.\ 2006.\ Terrestrial planet formation with 
strong dynamical friction.\ Icarus 184, 39-58. 

\item[--] O'Brien, D.~P., Walsh, 
K.~J., Morbidelli, A., Raymond, S.~N., Mandell, A.~M.\ 2014.\ Water 
delivery and giant impacts in the ``Grand Tack'' scenario.\ 
Icarus 239, 74-84. 

\item[--] Oka, A., Nakamoto, T., Ida, 
S.\ 2011.\ Evolution of Snow Line in Optically Thick Protoplanetary Disks: 
Effects of Water Ice Opacity and Dust Grain Size.\ The Astrophysical 
Journal 738, 141. 

\item[--] Ormel, C.~W., Klahr, H.~H.\ 2010.\ The effect of gas drag on the growth of protoplanets. Analytical expressions for the accretion of small bodies in laminar disks.\ Astronomy and Astrophysics 520, A43. 

\item[--] Paardekooper, S.-J., Mellema, G.\ 2006.\ Dust flow in gas disks in the presence of embedded planets.\ Astronomy and Astrophysics 453, 1129-1140. 

\item[--] Paardekooper, S.-J., Mellema, G.\ 2006b.\ Halting type I planet migration in non-isothermal disks.\ Astronomy and Astrophysics 459, L17-L20. 

\item[--] Paardekooper, 
S.-J., Baruteau, C., Crida, A., Kley, W.\ 2010.\ A torque formula for 
non-isothermal type I planetary migration - I. Unsaturated horseshoe drag.\ 
Monthly Notices of the Royal Astronomical Society 401, 1950-1964.

\item[--] Paardekooper, 
S.-J., Baruteau, C., Kley, W.\ 2011.\ A torque formula for non-isothermal 
Type I planetary migration - II. Effects of diffusion.\ Monthly Notices of 
the Royal Astronomical Society 410, 293-303. 

\item[--] Pan, L., Padoan, P., Scalo, 
J., Kritsuk, A.~G., Norman, M.~L.\ 2011.\ Turbulent Clustering of 
Protoplanetary Dust and Planetesimal Formation.\ The Astrophysical Journal 
740, 6. 

\item[--] Raymond, S.~N., Quinn, 
T., Lunine, J.~I.\ 2004.\ Making other earths: dynamical simulations of 
terrestrial planet formation and water delivery.\ Icarus 168, 1-17. 

\item[--] Raymond, S.~N., Quinn, 
T., Lunine, J.~I.\ 2006.\ High-resolution simulations of the final assembly 
of Earth-like planets I. Terrestrial accretion and dynamics.\ Icarus 183, 
265-282. 

\item[--] Raymond, S.~N., Quinn, 
T., Lunine, J.~I.\ 2007.\ High-Resolution Simulations of The Final Assembly 
of Earth-Like Planets. 2. Water Delivery And Planetary Habitability.\ 
Astrobiology 7, 66-84. 

\item[--] Robert, F.\ 2003.\ The D/H 
Ratio in Chondrites.\ Space Science Reviews 106, 87-101. 

\item[--] Rubie, D.~C., Jacobson, 
S.~A., Morbidelli, A., O'Brien, D.~P., Young, E.~D., de Vries, J., Nimmo, 
F., Palme, H., Frost, D.~J.\ 2015.\ Accretion and differentiation of the 
terrestrial planets with implications for the compositions of early-formed 
Solar System bodies and accretion of water.\ Icarus 248, 89-108. 

\item[--] Shakura, N.~I., Sunyaev, R.~A.\ 1973.\ Black holes in binary systems. Observational appearance..\ Astronomy and Astrophysics 24, 337-355. 

\item[--] Shu, F.~H., Shang, H., Lee, 
T.\ 1996.\ Toward an Astrophysical Theory of Chondrites.\ Science 271, 
1545-1552. 

\item[--]  Shu, F.~H., Shang, H., 
Glassgold, A.~E., Lee, T.\ 1997.\ X-rays and fluctuating X-winds from 
protostars..\ Science 277, 1475-1479. 

\item[--]  Shu, F.~H., Shang, H., 
Gounelle, M., Glassgold, A.~E., Lee, T.\ 2001.\ The Origin of Chondrules 
and Refractory Inclusions in Chondritic Meteorites.\ The Astrophysical 
Journal 548, 1029-1050.

\item[--] Staff, J., Koning, N., 
Ouyed, R., Pudritz, R.\ 2014.\ Three-dimensional simulations of MHD disk 
winds to hundred AU scale from the protostar.\ European Physical Journal 
Web of Conferences 64, 05006. 

\item[--] Takeuchi, T., Lin, 
D.~N.~C.\ 2002.\ Radial Flow of Dust Particles in Accretion Disks.\ The 
Astrophysical Journal 581, 1344-1355. 

\item[--] Tanaka, H., Takeuchi, 
T., Ward, W.~R.\ 2002.\ Three-Dimensional Interaction between a Planet and 
an Isothermal Gaseous Disk. I. Corotation and Lindblad Torques and Planet 
Migration.\ The Astrophysical Journal 565, 1257-1274. 

\item[--] Testi, L., and 10 
colleagues 2014.\ Dust Evolution in Protoplanetary Disks.\ Protostars and 
Planets VI 339-361. 

\item[--] Villeneuve, J., 
Chaussidon, M., Libourel, G.\ 2009.\ Homogeneous Distribution of $^{26}$Al 
in the Solar System from the Mg Isotopic Composition of Chondrules.\ 
Science 325, 985. 

\item[--] Volk, K., Gladman, 
B.\ 2015.\ Consolidating and Crushing Exoplanets: Did it happen here?.\ 
ArXiv e-prints arXiv:1502.06558. 

\item[--] Youdin, A.~N., 
Goodman, J.\ 2005.\ Streaming Instabilities in Protoplanetary Disks.\ The 
Astrophysical Journal 620, 459-469. 

\item[--] Walsh, K.~J., Morbidelli, 
A., Raymond, S.~N., O'Brien, D.~P., Mandell, A.~M.\ 2011.\ A low mass for 
Mars from Jupiter's early gas-driven migration.\ Nature 475, 206-209. 

\item[--] Warren, P.~H.\ 2011.\ Stable-isotopic anomalies and the accretionary assemblage of the Earth and Mars: A subordinate role for carbonaceous chondrites.\ Earth and Planetary Science Letters 311, 93-100.

\item[--]  Weidenschilling, 
S.~J.\ 1977.\ Aerodynamics of solid bodies in the solar nebula.\ Monthly 
Notices of the Royal Astronomical Society 180, 57-70. 

\item[--] Weidenschilling, S.~J.\ 1977b. The
 distribution of mass in the planetary system and solar nebula.\
 {\it Astrophysics and Space Science} 51, 153-158.

\item[--] Williams, J.~P., Cieza, L.~A.\ 2011.\ Protoplanetary Disks and Their Evolution.\ Annual Review of Astronomy and Astrophysics 49, 67-117. 

\end{itemize}
\end{document}